
\magnification=\magstep0
\tolerance=1000
\hsize=16.0truecm
\vsize=23.5truecm
\topskip=1truecm
\raggedbottom
\abovedisplayskip=3mm
\belowdisplayskip=3mm
\abovedisplayshortskip=0mm
\belowdisplayshortskip=2mm%
\def\al{$\alpha$}
\def\be{$\beta$}

\def\ls{\vskip 12pt}
\def\et{{\it et\thinspace al.}}
\def\mwc {MWC349}
\def\gapprox{$_>\atop{^\sim}$}
\def\lapprox{$_<\atop{^\sim}$}
\def\kms {km\thinspace s$^{-1}$}
\def\ref#1 {\noindent \hangindent=24.0pt \hangafter=1 {#1} \par}
\def\vol#1 {{\bf {#1}{\rm,}\ }}
\def\apj   {{\it Ap.~J.~{\rm,}\ }}
\def\apjl  {{\it Ap.~J.~(Letters){\rm,}\ }}

\def\aap   {{\it Astr.~Ap.{\rm,}\ }}

\def\ats   {{\it Astron.~Tsirk.~{\rm,}\ }}

%
%
%
%
%
%
\centerline{HYDROGEN MASERS. II: \mwc.}
\ls
\ls
\centerline{Vladimir S. Strelnitski $^{1,2}$,
Howard A. Smith $^2$, and Victor O. Ponomarev $^3$}
\vskip 48pt
\ls
\ls
\ls
$^1$ New Mexico Institute of Mining and Technology, Socorro, NM 87801

$^2$ Laboratory for Astrophysics, National Air and Space Museum, Smithsonian
Institution,

Washington, DC 20560

$^3$ P.N. Lebedev Physical Institute, 117924 Moscow, Leninsky
prospect 53, Russia
\vskip 96pt
\centerline{Submitted to the {\it Astrophysical Journal, Part 1}}
\vfill\eject
\centerline{ABSTRACT}
\vskip 6pt
The conditions of masing and lasing in the hydrogen recombination lines
(HLR) in the disk and outflow of \mwc\ are studied. Comparison of the
complete set of the observed \al-lines, from H2\al\ through H92\al,
with simple models of optically thin spontaneous emission shows that
observable HRL masing in this source is limited to the interval
of the principal quantum numbers $n \approx 10$ -- 36.

We use our analytical and numerical results for the conditions of
optimum masing (Strelnitski \et\ 1995b; ``Paper~2''),
and the morphological parameters of a photoevaporized
circumstellar disk modelled by Hollenbach \et\ (1994), to obtain
analytical approximations and quantitative estimates for the expected
unsaturated maser gain and the degree of saturation in HRL
masers in the {\it disk} of \mwc. It is shown
that the unsaturated maser gains of the IR and optical HRL
lasers should be very high but in fact all of these transitions are
strongly saturated. Due to saturation, their relative intensity
with respect to the spontaneous emission
should steeply drop toward smaller $n$'s which
explains why optical and near infrared HRL lasers are not seen
in \mwc. This result is quite general and makes the prognosis for
observable high-frequency lasers in other sources pessimistic as
well. An exception might arise due to special geometry,
if lasing were confined to very
small solid angles.

We show that weak masing from the {\it outflow} of \mwc\ is only
possible in low $n$ \al-lines ($n\,$\lapprox 20--30), and then only if the
outflow begins much closer to the central star than the Hollenbach \et\
model predicts ($\approx 3\cdot 10^{15}\,$cm). We therefore conclude
that maser
emission from the {\it disk}, rather than from the outflow, is
responsible for the observed weak amplification of the ''pedestal``
spectral components in the \al-lines with $n$\lapprox 40. The fluxes in
the lower frequency \al-lines are well explained by spontaneous emission
from the outflow, with proper corrections for free-free absorption.

We review the current state (and prospects) of observations of masing and
lasing hydrogen \be-lines in \mwc. We argue in favor of ``close in
$n$,'' rather than ``close in frequency'' \al/\be\ pairs for intensity
analysis, when maser amplification is present.
\vskip24pt
{\it Subject headings:} stars:early-type---stars: individual
(\mwc)---stars:mass loss ---masers---radio recombination lines
\vfill\eject
\centerline{1. INTRODUCTION}
\ls
\mwc, a well-known variable emission line star ($V\,1478~Cyg$) with
high IR excess and the brightest stellar radio continuum
source in our Galaxy, recently became the first known hydrogen
maser in space. Mart\'in--Pintado \et\ (1989a,b) detected strong,
double-peaked emission in its mm hydrogen recombination lines (HRL),
and even stronger submm double-peaked lines were subsequently
detected by Thum \et\ (1994a,b). Mart\'in--Pintado \et\ (1989a)
presented arguments for the maser nature of the detected lines and
demonstrated the principal possibility of high-gain masing in these
lines in the dense ionized {\it outflow} from the star, but failed to
explain the observed spectral pattern kinematically. Ponomarev \et\
(1989) attempted to explain this pattern, ascribing one of the
spectral peaks to the outflow and another one to the partly ionized,
infalling circumstellar disk. However, later,
with new observational evidence,
several authors (Gordon 1992; Planesas \et\ 1992; Thum \et\ 1992)
were able to successfully
ascribe {\it both} peaks to a circumstellar {\it Keplerian disk}.

The masing Keplerian disk model has further been elaborated upon by Thum
\et\ (1994a,b) and by Ponomarev \et\ (1994). The slight observed
variations of the 35\al--30\al\ line widths with intensity, and the
deep dip between the two spectral peaks of 30\al, were
respectively attributed by Thum \et\ (1992) and by Ponomarev \et\ (1994), to
the effects of {\it unsaturated} maser amplification.
Earlier, however, Ponomarev \et\ (1991) estimated that the 30\al\ line should
be at the threshold of saturation. Thum \et\ (1992; 1994a,b) found
several observational manifestations of increasing saturation toward
higher frequency submm lines and connected them with a decreasing role
of collisional relaxation with increasing line frequency. Saturation
seems to be a key issue for predicting HRL lasing in the IR and
optical; therefore it is one of the central topics of this paper and
of the previous one (Strelnitski \et\ 1995b; ``Paper~2'').

Originally the mm and submm HRLs in \mwc\ were decomposed
into a ``masing'' (double-peaked) component
associated with the disk, and a ``thermal'' (pedestal) component
attributed to the outflow (Thum \et\ 1992). However, several
recent papers claimed weak masing in
the pedestal component too (Mart\'in--Pintado \et\ 1994a,b; Thum \et\
1994b; 1995). In the present paper we attempt to develop
a self-consistent model of
HRL masing and lasing in the disk and outflow of \mwc, in order to
explain the full set of existing observations and to predict the properties
of yet unobserved, potentially lasing HRLs at higher frequencies.

We summarize, in
Section~2, the observed fluxes in most of the hydrogen \al-lines which
have so far been detected in this source --- from  H\al, at 6563 \AA,
through 92\al, at 3.6~cm. Comparison of the observed fluxes with
simple theoretical models for optically thin spontaneous emission
provides further evidence for masing, and delineates the limits of
``active'' frequencies.

Section~3 is devoted to the interpretation of the double-peaked maser
lines. Using the approach to saturation analysis developed
in Paper~2, we obtain a reliable {\it
observational} estimate of the degree of saturation for the 30\al\
maser, confirming the rough previous suggestions that this line
should be at the threshold of saturation. We use the morphological
parameters of the
photoevaporation disk model of Hollenbach \et\ (1994), and our
analytical and numerical results concerning maser gains in HRLs, to
calculate the expected values of the unsaturated gain and the degree of
saturation for \al-lines
in the interval 5\lapprox\thinspace $n\,$\lapprox\thinspace 35. We show that
all
the IR \al-line lasers
in \mwc\ should be strongly saturated, in accordance with the
saturation tendencies revealed by Thum \et\ (1994a,b) from their
observations of mm and submm masers. The expected
maser-to-spontaneous intensity ratios for these saturated lines
steeply decrease toward lower $n$, providing further support for our
previous conclusion (Strelnitski \et\ 1995a) that detectable
double-peaked \al-line lasing in \mwc\ should extend from the submm
into the IR domain --- and then vanish somewhere around $n \approx 10$. We
use the qualitative ideas about the interaction of saturated masing
lines developed in Paper~2 to explain the observed
``clustering'' of the mm and submm lines near a single value of
the separation of two spectral peaks in radial velocity.

An analysis of the conditions for masing in the \mwc\ outflow
(Section~4) brings us to the conclusion that although weak masing in
\al-lines is possible here for $n$\lapprox 30, the observed increase
of the pedestal relative intensity toward lower $n\,$ is more
probably due to amplification in the disk. The bulk of the higher
$n\,$ pedestal lines is fairly well explained by optically thin,
spontaneous emission from the outflow, when proper corrections for
the free-free absorption in the lines are included.

In Section~5 we review the observational and theoretical situation
of the masing hydrogen \be-lines.  We demonstrate that, in accordance
with our model predictions, the only known weakly masing
double-peaked \be-line (32\be) forms in the disk near the corresponding
\al-lines of
close $n$. We argue that, when masing is present, observations of
the ``close in $n$'' \al/\be\ pairs should be more instructive than
observations of ``close in frequency'' pairs. A prognosis for
detecting other \be-line masers in \mwc\ is given.

In Appendix~1, we argue that previous attempts to prove the
presence of masing in mm HRLs in \mwc\ are not fully satisfactory, and
we give a more rigorous proof using two observed parameters of 30\al\
--- the flux, and the interferometric angular scale of the emitting
medium --- and using only very general model considerations about maser
geometries in a Keplerian disk.

In Appendix~2, we derive a simple expression for the \be/\al\ line
spontaneous emissivity ratio. It can be useful for
estimating whether or not effects of optical depth (positive or
negative) are present in the observed \be/\al\ intensity ratio.
\ls
\ls
\centerline{2. SUMMARY OF THE HYDROGEN \al-LINE FLUXES IN \mwc}
\ls
In Figure~1 we plot the integrated
intensities of the most of hydrogen recombination \al-lines
observed so far in \mwc\ (we omitted some lines for the sake of the
figure's clarity).

The fluxes in clearly observed double-peaked,
narrow-band components are shown with dots; those in
single-peaked, broad-band  components (``pedestals'') --- with
squares. The latter symbol is
also used for poorly resolved profiles and for profiles with only
barely resolved double-peaked components. Filled symbols mark single
observations (when several are known, we normally choose the
brightest one); open symbols, averages for some period of
observations (taken from Thum \et\ 1992; 1994a). Three optical and
near IR lines (H\al, P\al\ and B\al) were corrected for
interstellar and circumstellar extinction using the extinction
curve given by Thompson \et\ (1977).

The two solid lines in Fig.~1 show the calculated spontaneous
emissivity ratios for the Baker and Menzel's ``Case~B'', for two
extreme density values poossible in the emission line region of
\mwc. The difference between the two lines is not significant. Since
these theoretical lines show only {\it ratios} of emissivities, their
position along the ordinate is arbitrary. The broken line gives the
spontaneous emissivity ratios, corrected for free-free absorption,
which is important for low-frequency ($n\,$\gapprox\thinspace 30) lines. The
correction procedure is described in Section~4. The position of this
theoretical line along the ordinate is also arbitrary.

Although the observational data in Fig.~1 are not
homogeneous and belong to (or are averaged over) different epochs,
we can clearly see some important tendencies.

The most striking fact is that theoretical curve for
optically thin spontaneous emission, with only the correction for
free-free absorption, describes satisfactorily the observed
fluxes in almost all the {\it pedestal} components, from the highest
frequency optical and near IR lines to the lowest frequency
radio lines --- more than 4 decades in frequency and almost 12 decades
in flux!

In contrast, the mm and submm double-peaked lines (dots in Fig.~1)
lie, as a group, considerably {\it higher} than the spontaneous
emission curve. This may be regarded as strong, though indirect,
evidence of their being {\it amplified}. The pedestal components of the
two submm lines, 26\al\ and 21\al, recently detected by Thum \et\
(1994b), also show evidence of amplification in Fig.~1.

To date, 10 double-peaked mm and submm hydrogen recombination lines
have been reported in MWC349 (Mart\'in--Pintado \et\ 1989a,b; Thum \et\
1994a,b; 1995): nine \al-lines --- 36\al, 35\al, 34\al, 31\al, 30\al,
29\al,
27\al, 26\al, 21\al, and one \be-line --- 32\be\ . The 36\al\ line is
the lowest frequency \al-line where the double-peaked
component first appears with intensity comparable to the pedestal
component. Both the peak flux density $S$ of the double-peaked
component and its integrated flux $F$ increase with line
frequency, attaining $S_{max} \approx 300\,$Jy and $F \approx
1\cdot10^{-20}\, {\rm W\,cm^{-2}}$ by 21\al\ (Thum \et\ 1994b).

The tantalizing question naturally arises: how far into the still
unexplored IR
domain does the phenomenon of double-peaked masing extend? There
is one evident limit: the well known optical and  near IR lines seem
to be satisfactorily described by optically thin, or slightly
thick, {\it spontaneous} emission (Fig.~1 and also Thompson \et\ 1977;
Hamann and Simon 1986). Smith \et\ (1995) and Strelnitski \et\ (1995a)
recently undertook to narrow the gap of
unobserved lines between submm and near IR. Smith \et\ detected
6\al\ (12.37
$\mu$m) and 7\al\ (19.06 $\mu$m) on the IRTF telescope,
and Strelnitski \et\ marginally detected 10\al\ (52.53 $\mu$m)
and obtained an
upper limit for 15\al\ (169.4 $\mu$m) with the KAO observatory.
Fig.~1 shows that whereas the
7\al\ line lies very close to the optically thin spontaneous curve, as
other optical and near IR lines do, the 10\al\ is $\approx\,$4
times brighter than the spontaneous curve predicts. The 15\al\ flux upper
limit lies about 10 times higher than the spontaneous curve at this
frequency.

Though Strelnitski \et\ (1995a) could not spectrally resolve 10\al\
to see whether or not it is double-peaked, they conjecture, on the
basis of the line's flux excess, that {\it observable lasing from
\mwc\ might extend into the far IR down to $n \approx 10$}. However,
at lower $n$ detectable lasing definitely seems to vanish.

An important characteristic of the masing mm and submm double-peaked
lines is their {\it relative} intensity with respect to the pedestal.
Along with the increase of the peaks' absolute intensity, the ratio
$I_c/I_{pk}\,$ of the pedestal's intensity at the central dip between
the peaks to the peaks' intensity decreases from the threshold 36\al\
line, where $I_c/I_{pk} \approx 1$, down to 30\al, where this ratio
falls to $\approx 0.07\,$ (Ponomarev \et\ 1994). It is seen from the
recent observations by Thum \et\ (1994b) that $I_c/I_{pk}\,$ is still
low at 26\al\ ($\approx 0.06$), but at 21\al\ it becomes much higher,
$>0.2$.

The following questions thus have to be answered:

(1) Why does the double-peaked maser first appear at 36\al?

(2) Why does its relative intensity with respect to the pedestal
component first increase and then decrease with decreasing $n$?

(3) Why do the observable masers, both the double-peaked and the pedestal
ones, eventually vanish with decreasing $n$?

We address the double-peaked and the pedestal components separately
in the two following sections.
\ls
\ls
\centerline{3. THE DOUBLE-PEAKED COMPONENT: DISK MASERS}
\ls
\centerline{{\it 3.1. Morphological Model}}
\ls
We analyse the conditions for HRL masing in \mwc\ in the
framework of the Hollenbach \et\ (1994) morphological model of a
circumstellar disk photoevaporized by a massive hot central star.
According to that model, in the case where the ``fast'' stellar wind
from the hot central star has normal parameters, photoevaporation of
the disk by $L_c\,$ photons from the star creates two distinct
regions:

(1) the inner region, $R \leq R_g = 10^{14} M_*/M_\odot\;$(cm), --- a
nearly static ionized atmosphere of the disk whose scale hight $H$
increases with the radial distance $R\ $ as
$$
H(R) = R_g\left({{R} \over {R_g}}\right)^{3/2} \eqno (3.1.1)
$$
and the base number density of electrons decreases with $R\,$ as
$$
N_{e0}(R) = 5.7\cdot 10^{29} \Phi^{1/2}_{49}R^{-3/2}\; {\rm cm^{-3}}
\eqno (3.1.2)
$$
in the bulk of the atmosphere, except in its innermost part where the
density increases toward the center somewhat faster ($\propto
R^{-9/4}$); $\Phi_{49}\,$ here is the star's Lyman continuum photon
luminosity, in $10^{49}\,$photon/s;

(2) the outer region, $R \geq R_g$ --- the ``disk wind,'' freely
outflowing ionized gas.
\par\noindent
We assume for \mwc-A: $M_* = 26 M_\odot$ (Ponomarev \et\ 1994), thus
$R_g \approx 3\cdot 10^{15}\,$cm. The quantity $\Phi_{49}\,$ is more
difficult to determinate for stars as peculiar as \mwc-A; therefore we
normalize Eq.~(3.1.2) on the 30\al\ line. We do so by adopting
$2R\,(30\alpha) = 1.2\cdot 10^{15}\,$cm (the interferometric
distance between the two hot spots, corresponding to the two spectral
peaks, Planesas \et\ 1992), and $N_{e0}\,(30\alpha)  = 4\cdot 10^7\,
{\rm cm^{-3}}$ (the optimum electron density for the amplification
of this line at $T_e = 10^4\,$K; Paper~2). This fixes $\Phi_{49}$
in Eq.~(3.1.2) at the realistic value of $\Phi_{49} \approx
1.1$. Thus, we adopt for the quasi-static part of the disk [dropping
the sub-index ``0,'' spreading thereby the $N_{e0}(R)$ value over
the thickness $H(R)$]:
$$
N_{e}(R) = 5.9\cdot 10^{29}\, R^{-3/2}\, {\rm cm^{-3}}\,. \eqno
(3.1.3)
$$
Combining this with the approximation for the optimum amplification
density derived in Paper~2 (good over the whole interval
5,000\thinspace K\lapprox\thinspace $T_e\,$\lapprox\thinspace
10,000\thinspace K):
$$
N_e^{max} \approx 8\cdot 10^{15}\, n^{-5.7}\;, \eqno (3.1.4)
$$
we obtain the following approximation for the radial distance
of the maximum unsaturated amplification:
$$
R_{max} \approx 1.8 \cdot 10^9\, n^{3.8}\,{\rm cm}\,.\eqno (3.1.5)
$$
The values of $R_{max}$ for the two extreme lines considered here,
35\al\ and 5\al, differ by about 3 orders of magnitude (Fig.~2). They are,
however, both located within the static atmosphere.
\ls
\ls
\centerline{{\it 3.2. Unsaturated Maser Gain}}
\ls
In the interval of $R_{max}$ considered here the maximum
amplification coherent path length
(with respect to the radial velocity --- the chord $l_{max}$
in the Keplerian disk; cf. Paper~1) --- is equal to $R_{max}$
to within a factor of order unity. Thus
the maximum unsaturated optical depth can be estimated as
$$
\tau^{max}_0 \approx k_{net}^{max}\,l_{max} \approx k_{net}^{max}\,
R_{max}\;. \eqno (3.2.1)
$$
Substituting into
Eq.~(3.2.1) $R_{max}$ from  Eq.~(3.1.5) and $k_{net}^{max}$ from
Eqs.~(3.1.4) and (3.1.5) of Paper~2, we find
$$
\eqalignno{\tau^{max}_0 &\approx - 2\cdot 10^2\,{{n}\over {10}}^{-4.2}
\qquad(T_e = 10^4\,{\rm
K})\;, &(3.2.2)\cr
\tau^{max}_0 &\approx - 1\cdot 10^3\, {{n}\over {10}}^{-4.7}\qquad
(T_e = 5\cdot 10^3\,{\rm K})\;.&(3.2.3)}
$$
These equations show, first of all, that there should be a {\it
threshold} principal quantum number (frequency) for \al-line masing
in the disk --- the highest value of $n \equiv n_{thresh}$ for which
$\tau^{max}_0$ first becomes \gapprox\thinspace 1.
Eq.~(3.2.2) gives, for example, $\tau^{max}_0 (36\alpha) \approx 0.9$,
and this
is in good agreement with the observed threshold of
the double-peaked masing at $n \approx 36$.
The cut-off of the masing at $n \approx 36$ is, in this model,
an ``optical'' effect based on the interplay of the density
dependence of maser gain with the density structure of the disk;
it does not require a {\it physical} cut-off of the disk as it does in
Thum's \et, 1994a, suggestion. We note, however, that in fact 36\al\
is masing in our model at $R \approx 1\cdot 10^{15}\,$cm --- not so
far from the edge of the static atmosphere ($R_g \approx 3\cdot
10^{15}\,$cm).

Eqs.~(3.2.2) and (3.2.3) show also that $\tau^{max}_0$ steeply increases
with decreasing $n$: from $\tau^{max}_0 \sim 1$ at $n \approx 35$,
through $\tau^{max}_0 \sim 10^1$ at $n \approx 20$, through
$\tau^{max}_0 \sim 10^2$ at $n \approx 10$, and up to $\tau^{max}_0
\sim 10^3$ at $n \approx 5$. Thus, if this photoevaporation disk
model for \mwc\ and the hydrogen population calculations
are basically correct,
{\it the unsaturated maser gains in the IR and optical hydrogen
recombination lines in} \mwc\ {\it should be very high}.

Why then are the observed fluxes in optical and near IR lines {\it
not} extraordinarily high and are, in fact, close to the optically
thin spontaneous theoretical curve, Fig.~1? To answer this question,
we examine more closely the conditions of {\it saturation} for the
disk masers.
\ls
\centerline{{\it 3.3. Saturation: Observational Evidence}}
\ls
Ponomarev \et\ (1991) briefly analysed the
conditions of saturation in the H30\al\ line in \mwc\
and showed that, with the observed flux and  the
geometrical scale of the emitting region estimated by Mart\'in--Pintado
\et\ (1989b) from temporal variations, this line is probably marginally
saturated.

A more detailed analysis of the connection between the observed line
parameters and the degree of maser saturation was given in Paper~2.
It was shown that the solid angle of maser emission --- the most
undetermined free parameter in the previous saturation analyses by
Thum \et\ (1992; 1994a) --- can actually be excluded from final
equations. Then
the geometry of the problem can be reduced to some {\it linear} scale
$l$, whose reliable estimate, for such a source as \mwc, can be
taken directly from interferometric observations. The degree of
saturation, as measured by the ratio of the actual average intensity
in the source $J$ to the ``saturation intensity'' $J_s$, was shown to
be
$$
{{J} \over {J_s}} \approx 0.2\, \biggl ({{D} \over {{\rm
kpc}}}\biggr )^2\, \biggl ({{n} \over {10}} \biggr)^{4.7}  \biggl ({{S}
\over {{\rm Jy}}}\biggr )\,\biggl ({{10^{13}\,{\rm cm}} \over
{l}}\biggr )^2\;, \eqno (3.3.1)
$$
where $D$ is the distance to the source, $S$ is the observed flux
density, and $l$ is the amplification path length.

We apply now this condition of saturation to the double-peaked
component of 30\al\ in \mwc, for which Planesas \et\ (1992) measured the
angular separation of the ``blue'' and ``red'' hot spots: $\theta_s =
0.065\,$ arcsec. It is convenient, in this case, to pass from $l$
in Eq.~(3.3.1) to $\theta_l = (l/D)$ --- the angular measure of
$l$ as it would be seen from Earth, had it been perpendicular to
the line of sight:
$$
{{J} \over {J_s}} \approx 1\cdot 10^{-7} \biggl ({{n} \over {10}}
\biggr)^{4.7}  \biggl ({{S} \over {{\rm Jy}}}\biggr )\,\biggl ({{{\rm
arcsec}} \over {\theta_l}}\biggr )^2\;. \eqno (3.3.2)
$$
To a factor close to unity the amplification chord length in a
Keplerian disk with our parameters is $l \approx s/2$, thus,
$\theta_l \approx 0.5 \theta_s \approx 0.03\; {\rm arcsec}$.
Substituting into Eq.~(3.3.2) this value and also $n = 30$,
$S_{30\alpha} \approx 25\,$Jy (Thum \et\ 1994a), we find: $J/J_s
\approx 0.5$. This confirms the suggestion made before by Ponomarev
\et\ (1991) and by Thum \et\ (1992; 1994a) that this line should be at
the threshold of saturation.

We emphasize that the forgoing analysis of 30\al\ saturation is based
on the directly measured quantities --- the flux density
and the angular scale --- and
is only weakly model dependent. Unfortunately, direct measurement of
the angular scale has only been done for the 30\al\ double-peaked
component so far, and we can not apply the above analysis to other
lines to see how the degree of saturation changes with $n$.
However, Thum \et\ (1992; 1994a,b) draw attention to some
observational indications of growing saturation toward higher
frequency lines: some broadening of the lines and levelling off of the
maser photon luminosity. In the framework of their saturation
analysis (which was reduced to the evaluation of the ``saturation
temperature'' $T_s\,$ --- an analogue of $J_s$) they find that maser
saturation ``becomes easier'' toward lower $n$ because $T_s\,$ for them
is smaller. These authors correctly notice, however, that such an
analysis is insufficient: only {\it comparison} of a saturation
parameter ($T_s\,$ or $J_s$) with the corresponding expected
parameters of radiation field in the source enables any
statement about the degree of saturation in a masing line.

In the theoretical analysis of saturation given in the next two
sections we shall show that the degree of saturation
indeed grows toward higher frequency HRL transitions in \mwc, and it
leads to important consequences concerning observability of HRL
lasers.
\ls
\ls
\centerline{{\it 3.4. Saturation: Theoretical Prognosis}}
\ls
Assuming the solid angle of maser radiation $\Omega$ to be constant
along the amplification path, and ignoring spontaneous emission with
respect to the maser emission in a strong maser, the {\it
unsaturated} growth of the average intensity with $|\tau_0|\,$ is given
by
$$
J_u \approx {{\Omega} \over {4\pi}}I_0 {\rm e}^{|\tau_0|}\,, \eqno
(3.4.1)
$$
where $I_0$ is the input intensity --- the intensity of the
continuum or spontaneous emission, whatever is brighter, at the
frequency of the line. In the Raleigh-Jeans domain (all the lines in
question fall into this domain) the input intensity due to continuum
is $2kT^c_b\,/\,\lambda^2 \le 2kT_e\,/\,\lambda^2$, where $T_b^c$
is the continuum brightness temperature and $\lambda_0\,$ is the
transition wavelength. It can also be shown (e.g.: Strelnitski 1974)
that the input intensity due to spontaneous emission is $2k|
T_x|/\,\lambda^2\,$ where $T_x\,$ is the transition excitation
temperature. Since $| \beta_{12}|\, \sim 1\,$ at the density of
maximum maser gain (Paper~2), we have $| T_x|\, \approx T_e\,/|
\beta_{12}|\, \sim T_e$. Considering that the exact value of $I_0$
is not very important when $|\tau|$ is high, we can thus assume
$$
I_0 \approx {{kT_e} \over {\lambda_0^2}}\,, \eqno (3.4.2)
$$

It was shown in Paper~2 that the saturation intensity for an HRL
maser can be represented as
$$
J_s \approx {{2h\nu_0^3} \over {c^2}}\,{{A_t + C_t + C_{21}} \over
{A_{21}}}\,, \eqno (3.4.3)
$$
where $A_t \equiv (A_{t1}^{-1} + A_{t2}^{-1})^{-1}\,$ and $C_t \equiv
(C_{t1}^{-1} + C_{t2}^{-1})^{-1}\,$ are the harmonic mean values of
the total Einstein $A\,$ coefficient and of the total collision rate
for the two maser levels, $A_{21}\,$ is the Einstein $A\,$
coefficient for the signal transition, and $C_{21}\,$  is the rate of
collisional relaxation between the maser levels. It was also argued
that, at the optimum density for maser amplification in a HRL,
$A_t\,$ in Eq.~(3.4.3) can always be ignored as compared with $(C_t +
C_{21})$ which, in its turn, can be approximated as
$$
\delta \equiv (C_t + C_{21}) \approx 3\cdot 10^{-9}\,n^5\,N_e\;.
\eqno (3.4.4)
$$
Substituting for $N_e$ the approximation (3.1.4) for $N_e^{max}$;
using: $A_{n+1,n} \approx 6.3\cdot 10^9\, n^{-5}\,$(s$^{-1})$ and
$\nu_0 = c/\lambda_0 \approx 6.6\cdot 10^{15}\,n^{-3}\,$(s$^{-1})$
(the common approximations valid for $n$\al-lines with $n \gg 1$);
and neglecting $A_t$ in Eq.~(3.4.3), as compared with $(C_t +
C_{21})$, we obtain from Eqs.~(3.4.1) -- (3.4.4), at a line's optimum
density:
$$
J_s \approx {{2h\nu_0^3} \over {c^2}}\,{{\delta} \over {A_{21}}}
\approx 2\cdot 10^{-2}\,n^{-4.7}\,, \eqno (3.4.5)
$$
and
$$
{{J_u} \over {J_s}} \approx 0.3\,n^{-1.3}\,\Omega\,{\rm e}^{|\tau_0|}
\;. \eqno (3.4.6)
$$
Note that Eq.~(3.4.5) is valid over the whole temperature range
considered here (5,000\thinspace K\lapprox\thinspace
$T_e\,$\lapprox\thinspace 10,000\thinspace K), whereas the factor
${\rm e}^{|\tau_0|}$ makes Eq.~(3.4.6) rather strongly temperature dependent.

The solid angle of maser radiation can be estimated in our disk model
of \mwc\ as
$$
\Omega \approx {{0.5 R_{max} H}\over {l_{max}^2}} \approx 0.5{{H}
\over {R_{max}}} \approx 0.9\cdot 10^{-8}\,R_{max}^{1/2} \approx 4
\cdot 10^{-4}\,n^{1.9}\,. \eqno (3.4.7)
$$
It is assumed in Eq.~(3.4.7) that the ``vertical'' size of the masing
hot spot is $\approx H$, and its ``horizontal'' size is $\approx 0.5
R_{max}$. The latter seems to be a reasonable estimate considering
the radial density scale in the model disk (limiting the ring of
maximum amplification), and the gradient of velocities in a Keplerian
disk, as is directly shown by the numerical modeling of maser
amplification in such a disk (Paper~1). The last two equalities in
Eq.~(3.4.7) are obtained by substituting $H\,$ from Eq.~(3.1.1), with
$R_g \approx 3\cdot 10^{15}\,$cm, by assuming $R = R_{max}$, and
by substituting Eq.~(3.1.4) for $R_{max}$.

Combining Eqs.~(3.4.6) and (3.4.7), we get, for the case of
\mwc:
$$
{{J_u} \over {J_s}} \approx 1\cdot 10^{-4}\,n^{0.6}\,{\rm e}^{|\tau_0|}
\;. \eqno (3.4.8)
$$
The exponential factor is decisive in Eq.~(3.4.8). If
the degree of saturation $J_u/J_s$ is indeed
$\sim 1$ at $n \approx 30$, where also $|\tau_0| \approx 6$ (both these
results seem to be reasonable based on comparisons of {\it observations}
with our models), then
$J_u/J_s$ should quickly become $\gg 1$ at smaller $n$'s, because of the
rapid increase of $|\tau|$ with decreasing $n$ [cf. Eqs.~(3.2.2-3)].

We therefore conclude that {\it all the hydrogen IR and
optical \al-line lasers in} \mwc\ {\it should
be deeply saturated}.
\ls
\ls
\centerline{{\it 3.5. Observability of IR and Optical Lasers}}
\ls
It is shown (e.g.: Strelnitski 1974) that the output intensity of a
saturated maser can be represented as
$$
I_s \approx {{4\pi} \over {\Omega}} J_s |\tau_0|\,, \eqno (3.5.1)
$$
where $\tau_0$ is the {\it unsaturated} optical depth. In the case
of a homogeneous amplification path of length $l$, $\tau_0\,$  for a
HRL is given by
$$
\tau_0 = (k_c' + k_l^0)\,l\;, \eqno (3.5.2)
$$
with $k_c'\,$ being the continuum absorption coefficient at the line
frequency and $k_l^0\,$ being the unsaturated line absorption
coefficient. However, for those HRL's that have been observed as strong
masers ($n\,$\lapprox\thinspace 35)  $k_c' \ll k_l^0$ (see Fig.~10 in
Paper~2), hence we can ignore $k_c'$ in Eq.~(3.5.2) as compared
with $k_l^0$ and write:
$$
\tau_0 \approx k_l^0\,l = {{h\nu_0 B_{12}} \over {4\pi \Delta
\nu}}\,N_1' b_1 \beta_{12}\,l\;, \eqno (3.5.3)
$$
where $B_{12}\,$ is the Einstein coefficient, $\Delta\nu\,$ is the
line width, $N_1'\,$ is the LTE population of level 1, and $b_1,\,
\beta_{12}\,$ are the usual departure coefficients.

Since the masing ``hot spots'' in \mwc\ are not spatially resolved as yet,
it is more convinient to discuss their observed radiation in terms of the
flux density rather than intensity, even in terms of the flux density
integrated over frequency --- for
easier comparison with spontaneous emission. The integrated flux density
due to the portion of the source
having a surface area $d\sigma$ and an outgoing intensity $I$
at the line center, is
$$
dF = I\,\Delta\nu\,d\sigma/D^2\;, \eqno (3.5.4)
$$
where $\Delta\nu$ is the line width, and $D$ is the distance to the source.

Substituting Eq.~(3.5.1) for $I$, Eq.~(3.4.5) for $J_s$ and Eq.~(3.5.3)
for $\tau_0$ into Eq.~(3.5.4), considering that $l\,d\sigma = dV_s$
is the elementary volume from which the saturated maser emission comes to
the observer, and integrating over the whole volume providing this emission,
we get:
$$
F_s = \int {dF_s} = {{h\nu} \over {D^2}}\,{{\delta} \over {\Omega}}\,
\int {N_1'b_1 \beta_{12} dV_s}\;. \eqno (3.5.5)
$$
In obtaining this equation we have also used the common relation between the
Einstein's $A$ and $B$ coefficients.

It is straightforward to obtain an analogous equation for the flux due
to optically thin spontaneous emission:
$$
F_{sp} = \int {dF_{sp}} = {{h\nu} \over {D^2}}\,{{A_{21}} \over {4\pi}}\,
\int {N_2'b_2 dV_{sp}}\;. \eqno (3.5.6)
$$

Deviding Eq.~(3.5.5) over Eq.~(3.5.6), we get:
$$
{{F_s} \over {F_{sp}}} = {{4\pi} \over {\Omega}}\,{{\delta} \over {A_{21}}}\,
{{\int {N_1'b_1 \beta_{12} dV_s}} \over {\int {N_2'b_2 dV_{sp}}}}\;.
\eqno (3.5.7)
$$
Of the three ratios composing the right-hand side of Eq.~(3.5.7) two are
obvious --- the ratio of the integrated numbers of emitting particles and the
ratio of the solid angles of emission ($4\pi$ for spontaneous emission and
$\Omega$ for maser emission). The third ratio, $\delta/A_{21}$ is the ratio of
the {\it rates} of emission, per particle per second. The rate of
the saturated maser emission is $\delta$ because it is determined by the
collisional ``recycling''of the primary radiative-radiative pumping (cf.
Paper~2). If the observed maser amplification occurs at the optimum gas
density, $N_e^{max}$, $\delta$ is obtained by combining
Eqs.~(3.4.4) and (3.1.4):
$$
\delta \approx 2\cdot 10^7\, n^{-0.7}\;. \eqno (3.5.8)
$$
Using the approximation for $A_{21}$ mentioned in Sect.~3.4, we find
$$
{{\delta} \over {A_{21}}} \approx 4 \cdot 10^{-3}\, n^{4.3}\;. \eqno (3.5.9)
$$

Eq.~(3.5.9) shows that the factor $\delta/A_{21}$ strongly
favors high $n$ transitions. The physical reason for this can be roughly
explained as follows. As is shown in Paper~2, $N_e^{max}$
is close to the density of thermalization, the density under
which the {\it net} rate of collisional
transitions across the levels in question, $\approx \delta\,(kT_e/h\nu_0)$,
surpasses the net rate of
spontaneous radiative transitions which is roughly $\sim A_{21}$.
Thus, close to the density of thermalization, $\delta/A_{21} \sim
(kT_e/h\nu_0) \propto n^3$. This explains, very roughly, of course, why
the $\delta/A_{21}$ factor in Eq.~(3.5.7) favors high $n$ levels.

The forgoing result is important for our prognosis of
the observability of HRL
lasers. If there is a gradient of density in the source, and hence
the spontaneous emission in a line comes not only from the region
favorable for lasing in this line, the ratio of the integrals
in Eq.~(3.5.7) can be $\ll 1$.
It is seen then from Eqs.~(3.5.7) and (3.5.9)
that {\it at small n saturated laser emission can not significantly
exceed spontaneous emission in principle}, unless the solid angle of
lasing is $\Omega \ll 1$. This greatly narrows the conditions of
laser detectibility: the source has either to be homogeneous in
density (this density being just optimum for lasing in a chosen line),
and/or the solid angle of the laser beam must be very small,
requiring a special source geometry. Of course, to be detected
in a small solid angle such a laser must also have a fortuitous
orientation with
respect to the observer. If these conditions are
not fulfilled, the laser emission will either be lost in a
bright spontaneous background or will miss the observer geometrically.

In the particular case of the model masing disk of \mwc, the solid angle
of masing, given by Eq.~(3.4.7), does decrease with decreasing $n$, but it
decreases slower than $\delta/A_{21}$ does [Eq.~(3.5.9)], thus, after
masers have become fully saturated, i.e. for $n\,$\lapprox\thinspace 20,
the contrast $F_s/F_{sp}$ should drop with decreasing $n$, and
at some value of $n$ the laser component should vanish
in the brighter background of spontaneous emission.
We can crudely estimate
this value of $n$ as follows. Taking as $F_s/F_{sp}\,$ for 21\al\
the ratio of the observed maser flux, shown in Fig.~1, to the level
of the spontaneous flux given in this figure by the solid theoretical
curve, $F_s/F_{sp} \approx 10$, we find, from $F_s/F_{sp} \propto
n^{2.4}\,$ [Eqs.~(3.5.7), (3.4.7), and (3.5.9)], that $F_s/F_{sp}$
drops to $\approx 1$
at $n \approx 8$. This agrees with the lack of a significant
lasing component at $n\,$\lapprox\thinspace 10 in Fig.~1.
\ls
\centerline{{\it 3.6. Saturation and the (Peak Separation) -- (Line
Frequency) Relation}}
\ls
Proper accounting for saturation may help solve one more puzzle
associated with the
multi-line HRL maser in \mwc. It was observed that the peak
separation $\delta v$ increased
with the decreasing $n$ in double-peaked masing $n\alpha$-lines ---
from $\approx 40\,$\kms\ at 36\al\ and 35\al\ to $\approx 50\,$\kms\
at 30\al\ (Thum \et\ 1992; 1995). This seemed to be in accord with the idea
that higher frequency lines should mase at higher density, thus
closer to the center of the Keplerian disk where the rotation
velocity is faster. However, attempts to explain the $\delta v
(n)\,$ correlation quantitatively, assuming a regular density
structure in the disk, met with difficulties. Ponomarev \et\ (1994)
argued that the observed ratio
$\delta v (30\alpha)/\delta v (35\alpha)\,$ would require an
improbably steep radial gradient of density (\al\ \gapprox\ 4 in $N_e
\propto r^{-\alpha}$), if each line mased at its density
of maximum unsaturated gain. The recently discovered 26\al\ and
21\al\ (Thum \et\ 1994a,b) are even closer in their peak separation
to each other and to 30\al\ (all have $\delta v \approx 50\,$\kms),
so that very unacceptable density gradients ($\alpha \sim
10-70\,$) are required to realize them at their densities of maximum
unsaturated gain in a Keplerian disk.

To explain the levelling off of $\delta v (n)$, Thum \et\ (1994b)
proposed that the disk rotation curve itself levels off inside some
radius. No physical mechanism for such levelling off in a
circumstellar disk is, however, proposed. Also improbable is
the hypothesis that the ionized disk ``finishes'' abruptly at $R \sim
40\,$A.U. from the star, where $N_e \sim 10^7 -10^8\,{\rm cm^{-3}}$,
so that all the submm lines are forced to form at this high-density
edge. At least, it is certainly not the case in the Hollenbach \et\
(1994) photoevaporation disk model we adopt here.

We hypothesize instead that the clustering of submm lines
in some ring of the
disk to produce close values of $\delta v$, is due to their
{\it interaction via saturation}. The forgoing analysis was based on
the assumption that masers and lasers in different lines are formed
independently from each other, at their densities of maximum {\it
unsaturated} amplification. Yet, only for the lowest $n$ lines are the
regions of maximum unsaturated amplification well separated in
the model disk of \mwc\ --- the higher the $n$, the more the
amplification regions for adjacent lines should overlap
(Fig.~2). We argued above that the disk masers and lasers with
$n$ \lapprox\ 30 are affected by saturation. A possible mechanism for
a maser to change its location under the influence of another
saturated maser was described in Paper~2. Saturation in a masing
line increases the decay rate of its upper level, thereby facilitating
energy sink for the adjacent line above. As a result, the
latter is able to mase at a higher density than its density of
maximum unsaturated masing, with the consequence that adjacent lines
will cluster spatially. One can suppose that this effect
in \mwc\ is initiated by
some lasing saturated IR transition below 21\al, at a density $\sim
10^8\, {\rm cm^{-3}}$, where the overlapping of negative
absorption coefficients for this group of transitions becomes large
enough. A verification of this mechanism by direct numerical
simulation is under way.
\ls
\ls
\centerline{4. THE PEDESTAL COMPONENT}
\ls
\centerline{{\it 4.1. Masing in the Outflow?}}
\ls
Some evidence for masing in the pedestal component of HRL, as opposed
to the double-peaked component, has appeared lately.

Mart\'in--Pintado \et\ (1994a) observed narrow, highly variable,
and therefore presumably masing features in 39\al\ and 35\al\ at
``non-double-peak'' radial velocities which they ascribe to the
outflow. In addition, Mart\'in--Pintado \et\ (1994b) argued that the observed
flux
in 66\al\ (which, like all the other lines with
$n\,$\gapprox\thinspace 36, has only
a pedestal component) was considerably higher than the expected
spontaneous emission, and hypothesized that it was due to maser
amplification.

Thum \et\ (1994b) find that the pedestal in 21\al\ is relatively
stronger than in 26\al\ and attributed this to effects of stimulated
emission in the outflow. Their observation of these two lines is
plotted in Fig.~1, and it is seen indeed that there is a significant
excess of the flux, relative to the probable optically thin LTE
value, for {\it both} the double-peaked {\it and} the pedestal
components, especially in the case of 21\al.

Thum \et\ (1995) find that the pedestal flux ratios of several
\be/\al\ pairs of nearly the same frequency --- from 26\al/33\be\ to
40\al/48\be\ --- are considerably smaller than the expected optically
thin LTE ratios, and explained the relative enhancement of the
\al-lines by the stimulated emission in the outflow.

In view of this observational evidence, it is interesting to
investigate the conditions of masing in the outflow of \mwc, as
compared with its disk. The radio continuum data for this source are
adequately explained by a model of spherically-symmetric expansion at
constant velocity (Olnon 1975). Adopting the value of the outflow
constant obtained by Cohen \et\ (1985), $N_e(R)\,R^2 = 0.81\cdot
10^{37}\, {\rm cm^{-1}}$ (which corresponds to the spherically
symmetric, 50 \kms\ constant velocity mass loss of $1.2\cdot
10^{-5}\,$M$_\odot/$yr), we will have for the radius $R(N_e)$ at
which the outflow has a given electron number density $N_e$:
$$
R(N_e) \approx 2.8\cdot 10^{18} N_e^{-0.5}\;. \eqno (4.1.1)
$$
The values of $k_{net}^{max}\,$ presented in Table~1 of Paper~2, were
calculated for the {\it thermal} line broadening alone, at the
temperature considered. To account for the line weakening due to the
expansion of the outflow, these values should be decreased by a
factor of $\approx w/\Delta v_D$, where $w\;$ is the observed line
width, due to expansion, and $\Delta v_D\;$ is the thermal Doppler
line width. Thus, in this case maser gain is
$$
|\tau^{max}| \approx |k_{net}^{max}|\, R(N_e^{max})\,(\Delta
v_D/w)\;. \eqno (4.1.2)
$$

If masing for each line occurs in its ``optimum'' layer
($N_e = N_e^{max}$), Eqs.~(4.1.1) and (4.1.2), and Table~1 of Paper~2,
with $w/\Delta v_D \approx 4$, yield the following estimates
for the maser gain at the lowest conceivable temperature, $T_e =
5,000\,$ K:~~ $|\tau|(21\alpha) = 7$; $|\tau|(30\alpha) = 1.8$;
$|\tau|(40\alpha) = 0.4$; $|\tau|(66\alpha) = 0.02$;
$|\tau|(92\alpha) = 0.002$. These values decrease
with increasing $T_e$:
at $T_e = 10,000\,$K, $|\tau|$ is $> 1\,$ for only
$n\,$\lapprox\thinspace 21.

These estimates show that maser gain $> 1\,$ in the outflow is
only possible for relatively low $n\,$ \al-lines,
$n\,$\lapprox\thinspace $30\,(\pm 10)$, with the exact values
depending on the temperature.

At longer wavelengths ($n\,$\gapprox\thinspace 30), Fig.~1 shows a gradually
increasing depression of the pedestal fluxes, as compared with the
``simple'' optically thin LTE model. This model (solid lines in
Fig.~1) assumes that radiation in all recombination lines comes from
the {\it same} volume of space. Then the flux is simply proportional
to spontaneous emissivity, which is $\propto n^{-6}\,$ for $n \gg 1$.
However, as can be seen in Fig.~10 of Paper~2, the free-free opacity
surpasses the line opacity at $n\,$\gapprox\thinspace 30,
and their ratio then
steeply increases with $n$ ($k_l'\,$ in this figure should be
reduced by the factor $w/\Delta v_D \approx 4-6\,$ to account
for the velocity dispersion in the outflow). Therefore, for the long
wavelength lines only emission from the volume located above the
surface $\tau_c \approx 1\,$ will be observed. In the approximation
of a spherically symmetric, constant velocity outflow with $N_0 R_0^2
= 0.81\cdot 10^{37}\, {\rm cm^{-1}}\,$ and $T_e = 10^4\,$K, the
radius at which $\tau_c = 1\,$ is
$$
R(\tau_c = 1) \equiv R_0 \approx 2.4\cdot 10^{16} \bigl ({{\nu} \over
{\rm GHz}}\bigr)^{-2/3} \, {\rm (cm)}\;, \eqno (4.1.3)
$$
and the volume emission measure above the level $\tau_c = 1$ is
$$
\int N_e^2\,dV = N_0 R_0^2\cdot 4\pi\int_{R_0}^\infty {dR \over R^2}
\approx 8.0\cdot 10^{74} R_0^{-1} \approx 3.4\cdot 10^{58} \bigl
({{\nu} \over {\rm GHz}}\bigr)^{2/3} \propto n^{-2}\;. \eqno (4.1.4)
$$

Thus, the luminosity and the flux in the lines with $n$ \gapprox 30
should be proportional to $n^{-6} n^{-2} \propto n^{-8}$. This
prediction is shown in Fig.~1 with a broken line. It
describes quite well the observed fluxes from the outflow for the whole
range $n \approx 30 - 92$. The greatest deviation from this
theoretical curve (a factor of $\approx 4$ excess) is the
observation of 66\al\ by Mart\'in-Pintado \et\ (1994b). These authors
concluded that maser amplification in 66\al\ (of about
this factor) takes place in the outflow and they support this
conclusion by model calculations, assuming a mass outflow rate of $9\cdot
10^{-5}\,$M$_{\odot}\,$yr$^{-1}$ --- 8 times higher than the value of
Cohen \et\ (1985) adopted here. With the latter, as argued
above, the negative optical depth of the outflow in 66\al\ is only
$\sim - 10^{-2}$, and masing with a gain $>1\,$ is impossible. The
66\al\ flux measured by Altenhoff \et\ is significantly lower and
closer to our theoretical curve. More observations in this line are
needed to clarify the possibility of time variations and masing.
\ls
\ls
\centerline{{\it 4.2. Masing Disk versus Masing Outflow}}
\ls
We showed in the preceding section that only HRL with $n$\lapprox
$30\pm 10\,$ can have maser gain \gapprox 1 in the model
spherically-symmetric outflow with the mass loss rate of \mwc.
However, lines with $n$\lapprox 40 require $N_e$\gapprox $5\cdot
10^6\,$cm$^{-3}$ for their masing (see Table~1 in Paper~2).
According to Eq.~(4.1.1), these high densities would occur in
the outflow only at $R\,$\lapprox $1\cdot 10^{15}\,$cm. Yet, in the
photoevaporation disk model adopted here, there is {\it no} outflow
so close to the star: the quasi-static disk's atmosphere extends up
to $R = R_g \approx 3\cdot 10^{15}\,$cm for a star of 26 M$_\odot\,$
(Sect.~3.1).

We conjecture therefore that the pedestal masing, especially masing
with a gain $> 1$, as in the 26\al\ and 21\al\ lines, occurs in the
{\it disk} of \mwc, rather than in its outflow.

The optical thickness of the disk increases with decreasing $n$. It
was demonstrated in Sect.~3.2 that a
dominant double-peaked component first appears at
$\approx$ 36\al\ where
gain in the chords can be $\approx 1$. But when this happens,
the lower $n$ lines in the outer parts of the disk already have their
gains {\it approaching unity} and can therefore undergo a weak
amplification. For example, at the physical edge of the static
atmosphere ($R = R_g$), where $N_e \approx 4\cdot 10^6\,$cm$^{-3}$,
the 40\al\ line would have $\tau \sim 1$, if its amplification path
length is $\sim R_g$; its weak amplification could thus produce
some structureless supplement to the spontaneous emission. Amplified
emission from the different radial velocities
within the disk (see Fig.~1 in Paper~1) which form the ``pedestal''
component should progressively increase with increasing $|\tau|$ and
decreasing $n$. This is
in accord with the decrease of the
(\be/\al)$^{spont}$/(\be/\al)$^{obs}\,$ ratio toward lower $n$, seen
in the Thum \et\ (1995) paper and suggestive of an increase in the
\al-line pedestal amplification with decreasing $n$.

At still lower $n$, not only the chords (which give the double-peaked
component) but also other parts of the disk acquire maser gain
sufficient for high gain masing, and they will form an ever growing
masing ``pedestal'' to the double-peaked component. This {\it
disk-originated} pedestal can be broad, because of the radial
velocity dispersion in the disk. The contrast between the double-peaked
component and the pedestal should decrease with
increasing saturation. Our numerical modeling of the masing in an
edge-on Keplerian ring (Paper~1) shows that in a fully saturated
regime the center-to-peak intensity ratio may increase up to
$I_c/I_{pk} \approx 0.5$. The increase of $I_c/I_{pk}\,$ from
$\approx 0.06\,$ by 26\al\ up to $\approx 0.2\,$ by 21\al, as
observed by Thum \et\ (1994b), is in good agreement with this model, given
the evidence of a growing degree of saturation with decreasing $n$. We
anticipate a further decrease of the peak-to-center intensity ratio
with decreasing $n\,$, due to the increasing saturation in the disk,
before spontaneous emission washes out both the double-peaked and the
``pedestal'' masing components, at $n\approx 10$.
\ls
\ls
\centerline{5. \be-LINES}
\ls
Five recombination \be-lines of hydrogen have been detected in \mwc\ so
far: 39\be\
(Gordon 1994), 48\be, 38\be, 33\be, and 32\be\ (Thum \et\ 1995).
Comparison of the intensities of \al/\be\ pairs proved to be an
effective tool in analyzing masing in \al-lines. Besides,
\be-lines themselves show some evidence
of increasing masing with increasing frequency,
and it appears likely a real high-gain maser in
the higher frequency \be-lines will be detected in the near future.
These two aspects of studying \be-lines in \mwc\ are discussed below.
\ls
\centerline{{\it 5.1. \be/\al\ Intensity Ratio}}
\ls
Gordon (1994) detected the first \be-line in MWC349 (39\be), and
he was the first to use a \be/\al\ intensity ratio (39\be/31\al, the
lines of approximately equal frequencies), as evidence of masing
--- in the 31\al\ {\it double-peaked} component.

Thum \et\ (1994c) used the \be/\al\ intensity ratios of several other
equal-frequency pairs they had detected, and demonstrated weak masing
in the \al-line {\it pedestals} (see Sect.~4).

The use of the \be/\al\ ratio is based on the assumption that both lines
of the pair are produced in approximately the same region of space,
and on the fact that $|k_{net}^{\alpha}|\,$ is
normally several times higher than $|k_{net}^{\beta}|$.
Because of the latter condition, the medium always becomes optically thick
in the \al-line first. Note that the relative effects of optical thickness
are just opposite when the
populations are inverted and when they are not.
In the former case the \be/\al\ intensity
ratio changes in favor of the \al-line, because of the higher maser
amplification. In the latter case
the ratio changes in favor of the
\be-line, as compared with the optically thin case, because of the
higher self-absorption in the \al-line.
Since in the \be/\al\ method the observed intensity
ratio is compared with the spontaneous emissivity ratio, it is useful
to have a simple approximation for the latter. We derive such an
approximation in Appendix~2.

For the two {\it pedestal} pairs, 39\be/40\al\ and 39\be/31\al, Gordon
(1994) measured intensity ratios of $\approx 0.6$ and 0.12,
respectively. Using, as the corresponding spontaneous emissivity
ratios, $\epsilon(39\beta)/\epsilon(40\alpha) \approx 0.6\,$
and $\epsilon(39\beta)/\epsilon(31\alpha) \approx 0.3$, he came to
the conclusion that these lines were observed in an
approximately LTE optically thin ratio. Our  Eq.~(A2.5) gives for the
$\epsilon(39\beta)/\epsilon(40\alpha) \approx 1.3\,$ --- twice as
much as the value adopted by Gordon. Comparison of our emissivity
ratios with the flux ratios observed by Gordon leads us to the
conclusion that both \al-line pedestals are amplified
by a small factor of $\approx 2$. We note a remarkably good agreement
of our $(\beta/\alpha)^{spont}/(\beta/\alpha)^{obs}\,$ ratio for the
39\be/40\al\ pair with the Thum's \et\ (1995) ratio for the
48\be/40\al\ pair: $\approx 0.5\,$ in both cases. This is another
confirmation that the 40\al\ pedestal is amplified by a factor of
$\approx 2\,$ (see Sect.~4.2).
\ls
\centerline{{\it 5.2. \be-Line Masing in the Disk}}
\ls
The highest frequency \be-line detected so far in \mwc,
32\be, is also the only one
demonstrating a {\it double-peaked} component along with the
broad pedestal component (Thum \et\ 1995). By analogy with the
double-peaked \al-lines, the double-peaked component of 32\be\ was
interpreted by Thum \et\ (1995) as a result of weak masing in the
disk. Here we develop this interpretation further.

First of all, using the rather precise information about the peaks'
{\it radial velocities} (Table~2 in Thum \et), we argue that the
32\be\ peakes were formed in the same region where the \al-line with
the closest observed $n\,$ (34\al) was formed, rather than in the
region where the 32\be\ {\it frequency} companion (26\al) was formed.
The red peak of 32\be\ ($33.6 \pm 2.3\,$\kms) is closer to that of
34\al\ ($32.3 \pm 1.0\,$\kms, as averaged for a long period of
observations; Thum \et\ 1994a) than to that of 26\al\ ($30.7 \pm
.1\,$\kms), though the difference (2-3 \kms) is comparable with the
observational errors or time variations (1-2 \kms). Yet, the blue
peak of 32\be\ ($- 13.3 \pm 1.5\,$\kms) is {\it much} closer to that
of 34\al\ ($- 13.0 \pm 1.5\,$\kms) than to that of 26\al\ ($- 21.1
\pm .1\,$\kms), and in this case the difference between the two
groups ($\approx 6\,$\kms) is significantly greater than the
uncertainties ($\pm 2\,$\kms).

If the two lines are indeedcoming from the same volume of space,
masing seems to be necessary to explain their observed flux ratio.
{}From Eq.~(A2.5) we have:$$
{{\epsilon_{32\beta}} \over {\epsilon_{34\alpha}}} \approx 1.6\;.
\eqno (5.2.1)
$$
Yet, the fluxes in the peaks of 32\be\ ($2 \pm 2\,{\rm Jy}\,$\kms)
and 34\al\ ($ 20 \pm 20\,{\rm Jy}\,$\kms) have a ratio of only
$\sim 0.1$, though with a high uncertainty. The theoretical optically
thin ratio (5.2.1) is thus much higher than the observed
ratio; this discrepancy is naturally explained by maser amplification
which is stronger in 34\al\ than in 32\be.

That masing \al-\ and \be-lines with close $n$ can indeed share the
same region of space, can be shown in the following general
manner. Every three adjacent levels, 1, 2, and 3 (Fig.~3) produce two
\al-lines ($2 - 1\,$ and $3 - 2$) and one \be-line ($3 - 2$). The
definition of excitation temperature [Eq.~(2.2.1) in Paper~2] and the
obvious identity $(N_1 g_2/N_2 g_1)\,(N_2 g_3/N_3 g_2)\,(N_3 g_1/N_1
g_3) = (n_1/n_2)\,(n_2/n_3)\,(n_3/n_1) = 1\,$ give the following
relation between the frequencies and excitation temperatures of the
three transitions (Strelnitski 1983):
$$
{{\nu_{31}} \over {T_{x31}}} = {{\nu_{21}} \over {T_{x21}}} +
{{\nu_{32}} \over {T_{x32}}}\;. \eqno (5.2.2)
$$
Since \be-coefficients of the adjacent \al-lines, and, therefore, the
excitation temperatures $T_{x21}\,$ and $T_{x32}$, are always close
to each other [see eqs.~(2.2.3) and (2.2.4) and Fig.~4 in Paper~2],
and since $\nu_{31} = \nu_{21} + \nu_{32}$, it follows from
Eq.~(5.2.2) that
$$
T_{x31} = T_{x21}\,{{\nu_{31}} \over {\nu_{21} + \nu_{32}\,{{T_{x21}}
\over {T_{x32}}}}}
$$
should also be close to $T_{x21}\,$ and $T_{x32}$. Thus, the
\be($N_e$) and the $k_l\,(N_e)\,$ profiles should be similar for the
\al- and \be-lines of close $n$. The $k_{net}\,(N_e)\,$ profiles will
also be similar for those lines with $k_l \gg k_c\,$ ($n$ \lapprox\
40; see Fig.~8 in Paper~2).

Fig.~6 in Paper~2 shows indeed that the $k_{net}\,(N_e)$ profiles are
quite similar for the \al- and \be-lines of close $n$. The
$k_{net}^{32\beta}/ k_{net}^{34\alpha}\,$ ratio, at the density of
their maximum gain ($N_e \approx 1.5\cdot 10^7\,{\rm cm^{-3}}$) is
$\approx 0.2$, at $T_e =\,$5,000 -- 10,000 K. Suppose,
$\tau_{net}^{34\alpha} \approx - 3$. Then $\tau_{net}^{32\beta}
\approx - 0.6$, and their intensity ratio, produced by unsaturated
amplification, will be $\approx {\rm e}^{\Delta\tau} \approx 10$, as
observed. According to Fig.~5, Paper~2, to produce these optical
depths at $10,000 >T_e >\,$5,000 K an amplification length of
50 -- 100 A.U. would be sufficient, again in accordance with the
interferometric distance of 80 A.U. between the 30\al\ hot spots
(Planesas \et\ 1992).

If the model of a masing Keplerian disk, with density increasing
toward the center, is correct, then 32\be\ should be the
``threshold'' of masing
for \be-lines, as 36\al\ is for \al-lines (Thum \et\ 1995). The
value of the maximum $|k_{net}|$ is about an order of magnitude
smaller for
a \be-line than for an \al-line of close $n$, which is why at the
radial distance where the ``threshold'' \al-line is formed the
corresponding \be-lines are still optically thin and thus not
amplified enough. The threshold \be-line, with $|\tau_{net}| \approx
1$, should appear at a smaller radial distance (higher density),
where the corresponding \al-line is already optically thick. This
explains why the threshold of double-peaked masing occurs at smaller
$n\,$ for \be-lines (32\be) than for \al-lines (36\al).

The intensity of double-peaked masing in \be-lines should
increase further with decreasing $n$, as happens among \al-lines. Using
Fig.~6$g$ of Paper~2 and Eq.~(3.1.5) of this paper, we find, for
example, that the maser gains of the 29\be\ and 26\be\ lines (both
having good conditions for ground-based observations) should be,
respectively, $\approx 1.5\,$ and 2.5
times higher than the gain in 32\be. Hence, maser amplification
several times stronger
in 29\be\ and 26\be\ lines is anticipated. Detection of a masing
double-peaked component in these lines, which have very close
observed \al-companions (29\al---30\al\ and 26\al---27\al,
respectively), will be especially useful for a
comparative analysis similar to that presented
for the 32\be/34\al\ pair in Thum \et\ (1995)
and in this section.

Suborbital and orbital IR observatories have a good chance of
detecting, in the near future, even stronger lasers in higher
frequency \be-lines --- from $\approx\,$ 20\be\ through $\approx\,$
15\be\ ($\approx 200 - 100 \mu$m), where the unsaturated maser gains
can be considerably higher than unity.

In conclusion, we emphasize once more that an important difference
exists between the traditional analysis of {\it optically thin}
\al/\be\ pairs and the new situation where {\it masing} is
responsible for the observed intensities: a ``close frequency''pair
is preferable in the first case, while a ``close $n$'' pair --- in
the second.
\ls
\ls
\centerline{6. CONCLUSION}
\ls
\mwc\ is the only source in which strong HRL masers have been detected;
moreover,
it is a stellar source in which most HRL \al-transitions (both masing
and non-masing) have been observed ---
more than twenty! This provides us a unique opportunity to study in detail
the conditions needed for high gain HRL masing in the larger general context
of HRL emission. Such a
study was undertaken in this paper and Paper~2, and can be summarized as
follows:

(1) The factor $N_e^2\,$ plays a decisive role in determining the
gain $|\tau|$ of an HRL maser.

(2) At high principal quantum numbers $n$, the practical creation of a high
gain
HRL maser is hindered by the large dimensions needed to make $|\tau|
> 1$, a requirement caused by the low densities involved in inverting
high $n\,$ transitions, and by free-free absorption which reduces the
gain.

(3) Low $n$ transitions require high densities for inversion,
and if present in a source (as is the case for \mwc), the unsaturated gain
may become very large even with small dimensions. However, masers or
lasers at low $n$ saturate easily, and tend to disappear in the
background of the spontaneous emission of the source unless the
solid angle of the maser beam is very small.

(4) In the case of \mwc\ the optimum interval for strong \al-line
masing and lasing lies between these two extremes,
10\thinspace\lapprox\thinspace $n\,$\lapprox\thinspace 36.
{\it Relatively strong, double-peaked IR
\al-line lasers may, therefore, be discovered in this source in the
interval  $n \approx 10-20$}. Their expected fluxes can be estimated
from Fig.~1.

(5) A $|\tau| > 1$ \al-line maser in the {\it outflow} of \mwc\
doesn't seem possible for $n\,$\gapprox\thinspace 30--40.

(6) Fig.~1 confirms the hypothesis of Thum \et\ (1994b) that the
relative increase of the pedestal component in 26\al\ and 21\al\ is
due to masing. But we show the observed pedestal masing probably occurs in
the {\it disk}, not in the outflow. We predict further increase
of the pedestal-to-peak flux ratio toward lower $n$, until
at $n \approx 10$ the whole
structure is lost in the steeply growing spontaneous emission
of the source.

(7) High-gain, double-peaked \be-line masers and lasers with $n <
32$ should be present in the submm and IR domain, with 29\be\ (614
$\mu$m) and 26\be\ (447.4 $\mu$m) being the best candidates for the
ground-based searches.

(8) \be/\al\ line pairs of close $n$ transitions are more informative
in the case of
masing than the close-frequency line pairs.
\vfill\eject
\centerline {APPENDIX 1:~~{\it Prof of the Need for Masing}}
\ls
Though the hypothesis of maser amplification in the mm and submm HRL
in \mwc\ is now widely accepted, paradoxically enough no strict
proof for this has yet been given. In this Appendix we briefly
review the existing attempts to prove maser action occurs in these
lines. We demonstrate their insufficiency, and propose a more rigorous
proof.

The high intensities of the double-peaked lines ---
higher than one would expect
if the emission were purely spontaneous --- was the primary
reason to suspect maser amplification (Mart\'in--Pintado \et\ 1989a).
These authors demonstrated, by model calculations, that the observed
intensity of 30\al, for example, could not be produced by
thermalized emission from the expanding envelope of \mwc, the location
where the
lower frequency recombination radio lines were thought to arise.
However, it is known now that the double-peaked lines do {\it not}
originate in this outflow, so this modeling can no longer be taken as
a proof of their masing.

Mart\'in--Pintado \et\ (1989b) made an attempt to guess
the {\it size} of the emitting ``hot spots'', with the (rather
arbitrary) assumption that the upper limit of the size is defined
by the observed time scale of the intensity variations and the
maximum observed velocity of the gas motions around the star. They
obtained a size of \lapprox$\,4\cdot10^{14}\,$cm which, at the
distance of 1.2 kpc (Cohen \et\ 1985), corresponds to an angular
size of \lapprox$\,0.022\,$arcsec. The
observed flux densities then give a brightness temperature in the hot
spot of $T_b$ \gapprox$\, 1 \cdot 10^6\ $K --- implying some
amplification, because the highest conceivable brightness
temperature of the input radiation for the maser is less then $10^4 - 10^5\,$K.

Two more arguments were put forward in support of masing in these
double-peaked lines: their time variability (Mart\'in--Pintado \et\
1989b; Thum \et\ 1992), and
the narrowness of their spectral profile (Thum \et\ 1992; Ponomarev
\et\ 1992). Both phenomena are more easily explained with the hypothesis
of unsaturated maser amplification.

The last two arguments are, however, indirect, while the first two are
based, respectively, on the incorrect association of these lines with the
outflowing gas,  and on a mere guess of the radiating region's size.
Thus none of them can be taken as a strict proof of masing. Yet,
the direct measurement of the angular separation between the
two hot spots responsible for two spectral peaks in 30\al\
(Planesas, Mart\'in--Pintado, and Serabyn 1992) actually enable us
to prove masing quite rigorously, as follows.

According to Planesas \et\ (1992) the
angular distance between the centers of the two hot spots is
$\theta_s = (0.065 \pm 0.005)$ arcsec $= (3.2 \pm
0.2)\,10^{-7}\,$rad.  This separation can serve
as a strict upper limit to an individual hot spot's angular
diameter, $\theta_d$. If the separation between the two spectral
peaks is due to the motions of the gas (which seems very probable),
there should be some
connection between the radial velocity scales and the geometrical
scales. In
particular, the narrowness of the peak's spectral width, $\Delta v$,
as
compared with the separation between the peaks, $\delta v$, ($\Delta v
: \delta v \approx 1:4$), is an indication that the ratio of the
linear diameter of a hot spot to the distance between the hot
spots, $d:s = \theta_d : \theta_s$, should also be about $1:4$.
Our numerical modeling of maser amplification in an
edge-on Keplerian disk (Ponomarev \et\ 1994)
indeed gives $d:s$ \lapprox\ 0.3. Thus, we can safely admit that
at least $d^2 \ll s^2$ (or $\theta_d^2 \ll \theta_s^2$) is true.

Suppose that the observed flux $F$ within
one
peak of the 30\al\ double-peaked profile were due to isotropic,
optically thin spontaneous emission. In this case
$$
F = {{L_{sp}} \over {4\pi D^2}}\;, \eqno(A1.1)
$$
\noindent
where $D$ is the distance to the source;
$$
L_{sp} = N_2A_{21} h\nu_0 V \eqno (A1.2)
$$
\noindent
is the total spontaneous luminosity of the medium volume $V$
with the upper state population $N_2$;
$A_{21}$ is the Einstein coefficient for spontaneous emission,
$\nu_0$ is the frequency of the
transition, and $h$ is the Planck constant.
Unless the geometry of a hot spot is very unusual, its ``vertical''
dimension should also be $\approx d\,$ (or \lapprox $d$, as it can be in
the case of an edge-on disk). Then
$$
V \leq d^2 l \ll s^2 l,~~ {\rm or}~~ V \ll (\theta_s D)^2 l\;,
\eqno (A1.3)
$$
\noindent
where $l$ is the extension of the emitting region along the line of
sight.
Combining equations (A1.1) -- (A1.3) leads to the following strict
limit for the integrated flux due to spontaneous emission:
$$F \ll  {{A_{ji}} \over {4\pi}} \theta_s^2 N_j\, l\, h\nu_0
\;.\eqno (A1.4)$$
The population density of a hydrogen atom level with principal
quantum number $n$ is given by the Saha-Boltzmann equation:
$$
N_n = {\left({{h^2} \over {2\pi mkT_e}}\right)}^{3/2}N_e^2 n^2 {\rm
e}^{{I_n}
\over {kT_e}} b_n\;, \eqno (A1.5)
$$
\noindent
where $N_e$ is the electron number density; $I_n$ is the ionization
energy of the state $n$; $b_n$ is the departure coefficient; $k\,$ is
the
Boltzmann constant, and $m\,$ is the electron mass. For $n
\approx 30$, $b_n$ is
close to one if $N_e$ \gapprox $10^6\,$cm$^{-3}$, and for the
temperatures considered here,
$3000 {\rm K} \le T_e \le 10000\,{\rm K}\,$ the factor e$^{{I_n}
\over {kT_e}}$ is very close to one too. Thus, we have for the $n =
30$ level:
$$N_n \approx 4 \cdot 10^{-19} N_e^2\;. \eqno (A1.6)$$
\noindent
Substituting into Eq.~(A1.4): $A_{31,30}\approx 225\, {\rm s}^{-1}$,
together with the Eq.~(A1.6) and with the observed value
$F_{30\alpha} \approx
1 \cdot
10^{-14} {\rm {{erg} \over {cm^2 s}}}$ (Thum \et\ 1992), we obtain
finally the following strong requirement for the emission measure:
$$N_e^2 l \gg 1 \cdot 10^{31}\; {\rm cm^{-5}}\;. \eqno (A1.7)$$
For a Keplerian disk $l$\lapprox $s\,$ (Ponomarev \et\ 1994). With
the above interferometric value of
$\theta_s$ and with the distance to the source $D = 1.2\,$kpc, we
have
$l$\lapprox $s = \theta_s D = 1.2 \cdot 10^{15}\, {\rm cm}$. Then
another strong requirement, for the electron density, follows
from Eq.~(A1.7):
$$
N_e \gg 9 \cdot 10^7\, {\rm cm^{-3}}\;. \eqno (A1.8)
$$
We emphasize that (A1.7) and (A1.8) are {\it strong}
inequalities, becose of the probable small size of $d$,
as compared with $s$. But even these lower limits for $N_e^2 l$ and
$N_e$ are high enough to
show
the model of optically thin spontaneous emission
is internally contradictory.

Indeed, if the 30\al\ line were due to optically thin spontaneous
emission alone, the observed line width, $\Delta
v \approx 12\,{\rm km\,s,^{-1}}$  corresponds to a kinetic
temperature $T_e \approx 3000\;$K. The Rayleigh-Jeans approximation
is valid for mm lines, and the optical depth in a line of
spectral width $\Delta\nu$ is given by (Lang 1974):
$$
\tau_{ij} \approx {{\pi e^2} \over {mc}} f_{ij} {{N_i} \over
{\Delta\nu}} {{h\nu} \over {kT_x}} b_n l\;, \eqno (A1.9)
$$
\noindent
where $T_x$ is the excitation temperature of the transition.
At the densities
set by Eq.(A1.8) the 30\al\ transition should be deeply
thermalized (see Fig.~6$b$ in Paper~2). With $\Delta\nu \approx \nu_0 \cdot
{{\Delta
v} \over {c}} \approx 9\cdot 10^6\;$Hz (and, correspondingly,
$T_x = T_e \approx 3000\;$K), and with the oscillator strength for
hydrogen \al-transitions being
$f^{\alpha}_n \approx 0.194 n$ (Menzel 1969), we have
$$
\tau_{ij} \approx 8 \cdot 10^{-31} N_e^2 l\;. \eqno (A1.10)
$$
\noindent
Substituting (A1.7) for $N_e^2 l$ gives $\tau \gg 1$ which
contradicts our assumption of optical thinness; thus this
explanation of the observed value of $F\,(30\alpha)$ is contradicted.

Let us suppose instead that the 30\al\ line is thermalized and optically
thick. In this case the flux density in the line center is
$$
S_{max} \approx B(\nu_0, T_x) \cdot \theta_d^2 \ll B(\nu_0, T_x)
\cdot \theta_s^2\;,
\eqno (A1.11)
$$
where $B(\nu_0, T_x)$ is the Planck function. Substituting into
Eq.~(A1.11):
$B(232\,{\rm GHz}, 3000\,{\rm K}) \approx 5 \cdot 10^{-11}~{\rm
erg/(cm^2\,s\,Hz\,sr)}$,
and $\theta_s^2 \approx 1.0 \cdot 10^{-13}~{\rm sr}$, we obtain, as
an upper
limit for the possible flux density: $S_{max} \ll 0.5~$Jy, much
less than the observed value, $S \approx 25~$Jy.

We conclude that neither optically thin nor optically thick
spontaneous emission alone can explain the observed flux density
and integrated flux in 30\al, and, therefore, maser amplification of
the line in an optically thick inverted medium is necessary.
\vfill\eject
\centerline{APPENDIX 2: {\it Spontaneous Emissivity Ratios for
\be/\al\ Line Pairs}}
\ls
The integrated spontaneous emissivity in the line, whose
lower and
upper states' principal quantum numbers are, respectively,
$l$ and $u$, is
$$
\epsilon _{ul} = {{h\nu_0} \over {4\pi}}N_u A_{ul}\;, \eqno(A2.1)
$$
where $h$ is the Planck constant, $\nu_0\,$ is the transition
frequency, $N_u$ is the upper state level population,
$$
A_{ul} = 7.5 \cdot 10^{-22}{\nu_0}^2 f_{lu} \biggl ({l \over
u}\biggr)^2 \eqno (A2.2)
$$
is the Einstein coefficient, and $f_{lu}$ is the oscillator
strength.

The transition frequency is given approximately by
$2cR/l_{\alpha}^3$ for
$\alpha$-transitions and $4cR/l_{\beta}^3$ for $\beta$-transitions,
with $R$ being the Rydberg constant (Lang 1974). Thus
$$
{{\nu_{0\beta}} \over {\nu_{0\alpha}}} \approx
2\biggl({{l_{\alpha}}
\over {l_{\beta}}}\biggr)^3\,.\eqno(A2.3)
$$
We emphasize that the values of $l$ and $u$ for the \al~transition
may differ from those of the \be-transition in an \al/\be\ pair.

When the upper levels of the \al- and \be-transitions
are different, but not too distant, we obtain
from Eq.~(A1.5), ignoring the small difference in the factor
$e^{{I_n}\over {kT_e}}$:
$$
{{N_{u_{\beta}}} \over {N_{u_{\alpha}}}} \approx \biggl({{u_\beta}
\over {u_\alpha}}\biggr)^2\, {{b_{n(u\beta)}} \over
{b_{n(u\alpha)}}}\,. \eqno (A2.4)
$$
With $f_{lu} \approx 0.194l$ for \al-lines, $f_{lu} \approx 0.027l$
for \be-lines (Menzel, 1969), we
have finally from Eqs. (A2.1) through (A2.4):
$$
{{\epsilon_{\beta}} \over {\epsilon_{\alpha}}} \approx
1.1 \biggl({{l_{\alpha}} \over
{l_{\beta}}}\biggr)^6\,{{b_{n(u\beta)}} \over {b_{n(u\alpha)}}}
\,.\eqno (A2.5)
$$
If both $l_{\alpha}$ and
$l_{\beta}$ are $\gg 1$ and $|l_{\alpha} - l_{\beta}|$ \lapprox\
10, the uncertainty of this approximation is \lapprox 10\%, even if
the $b_n$ ratio is set equal to unity.

For the particular case of a pair of
\al-\ and \be-lines of close frequencies, the condition
$\nu_{0\alpha}
\approx \nu_{0\beta}$, combined with Eq. (A2.3), implies
$l_{\beta}/l_{\alpha} \approx 2^{1/3}$, and we get from Eq. (A2.5):
$$
{{\epsilon_{\beta}} \over {\epsilon_{\alpha}}} \approx
0.3 \qquad (\nu_{0\beta} \approx \nu_{0\alpha})\;.\eqno (A2.6)
$$
\ls
\ls
V.P. thanks the partial support of this research by the Tomalla
Foundation Fellowship and the International Center for   Fundamental
Physics in Moscow for its help in obtaining this   Fellowship. VSS and VOP
thank the Smithsonian Institution, National Air and Space Museum for   a
senior fellowship and
visiting fellowship, respectively, to work in the Laboratory for Astrophysics
on this program and
related observations.  Together with HAS they acknowledge financial
support from the
Institution's Scholarly Studies Program.  HAS also acknowledges partial
support from NASA grant NAGW-1711.
\vfill\eject
\centerline{REFERENCES}
\vskip 6pt
\ref{Baker, J.G., and Menzel, D.H. 1938, \apj\vol{88} 52.}
\ref{Brugel, E.W., and Wallerstein, G. 1979, \apjl\vol{229} L23.}
\ref{Cohen, M., Bieging, J.H., Dreher, J.W., and Welch, W.J. 1985,
\apj\vol{292} 249.}
\ref{Goldberg, L. 1966, \apj\vol{144} 1225.}
\ref{Gordon, M.A. 1992, \apj\vol{387} 701.}
\ref{Gordon, M.A. 1994, \apj\vol{421} 314.}
\ref{Greenstein, J. 1973, \apj\vol{184} L23.}
\ref{Hamann, F., and Simon, M. 1986, \apj\vol{311} 909.}
\ref{Hartmann, L., Jaffe, D., and Huchra, J.P. 1980, \apj\vol{239}
905.}
\ref{Hollenbach, D., Johnstone, D., Lizano, S., and Shu, F. 1994,
\apj\vol{428} 654.}
\ref{Lang, K.R. 1974, Astrophysical Formulae, Springer Verlag.}
\ref{Litvak, M.M. 1968, in ``Interstellar Ionized Hydrogen,'' ed.:
N.Y.~Terzian, N.Y. -- Amsterdam, Benjamin, p.713.}
\ref{Mart\'in--Pintado, J., Bachiller, R., Thum, C., and Walmsley,
C.M. 1989a, \aap\vol{215} L13.}
\ref{Mart\'in--Pintado, J., Thum, C., Bachiller, R. 1989b,
\aap\vol{229} L9.}
\ref{Mart\'in--Pintado, J., Neri, R., Thum, C., Planesas, P., and
Bachiller, R.
1994a, \aap\vol{286} 890.}
\ref{Mart\'in--Pintado, J., Gaume, R., Bachiller, R., Johnston, K.,
and Planesas, P. 1994b, \apjl\vol{418} L79.}
\ref{Olnon, F.M. 1975, \aap\vol{39} 217.}
\ref{Planesas, P., Mart\'in--Pintado, J., and Serabyn, E. 1992,
\apjl\vol{386} L23.}
\ref{Ponomarev, V.O., Smirnov, G.T., Strelnitski, V.S., Chugai, N.N.
1989, \ats\vol{1540} 5.}
\ref{Ponomarev, V.O., Strelnitski, V.S., Chugai, N.N. 1991,
\ats\vol{1545} 37.}
\ref{Ponomarev, V.O., Smith, H.A., and Strelnitski, V.S. 1994,
\apj\vol{424} 976 (Paper~1).}
\ref{Smith, H.A., Strelnitski, V.S., Kelley, D., Lacy, J., and Miles,
J. 1995, in preparation.}
\ref{Strelnitski, V.S. 1974, {\it Soviet Physics
Uspekhi~{\rm,}\ }\vol{17} 507.}
\ref{Strelnitski, V.S. 1983, {\it Nauchnye Informatsii Astron.
Council USSR ac. Sci.~{\rm,}\ }\vol{52} 75. (Spanish translation:
{\it Ciencia}, \vol {43} 185.)}
\ref{Strelnitski, V.S., Smith,  H.A., Haas,  M.R., Colgan, S.W.J.,
Erickson,  E.F., Geis, N., Hollenbach, D.J., and Townes, C.H. 1995a,
In: Proceedings of the Airborne Astronomy Symposium on the Galactic
Ecosystem: From  Gas to Stars to Dust, ed. M.R. Haas, J.A. Davidson,
and E.F. Erickson; San-Francisco: ASP, 1995, p.271.}
\ref{Strelnitski, V.S., Ponomarev, V.O., and Smith,  H.A. 1995b, \apj
submitted (Paper~2).}
\ref{Thompson, R.I., Strittmatter, P.A., Erickson, E.F., Witteborn,
F.L., and Strecker, D.W. 1977, \apj\vol{218} 170.}
\ref{Thum, C., Mart\'in--Pintado, J., and Bachiller, R. 1992,
\aap\vol{256} 507.}
\ref{Thum, C., Matthews, H.E., Mart\'in--Pintado, J., Serabyn, E.,
Planesas, P., and Bachiller, R. 1994a, \aap\vol{283} 582.}
\ref{Thum, C., Matthews, H.E., Harris, A.I., Tacconi, L.J.,
Shuster, K.F., and Mart\'in--Pintado, J. 1994b, \aap\vol{288} L25.}
\ref{Thum, C., Strelnitski, V.S., Mart\'in--Pintado, J., Matthews,
H.E., and Smith, H.A. 1995, \aap, submitted.}
\vfill\eject
\centerline{FIGURE CAPTIONS}
\ls
Fig.~1. Observed fluxes in hydrogen recombination \al-lines in \mwc.
{\it Dots}: the double-peaked component. {\it Squares}: the pedestal
component, or an unresolved line.  Filled symbols --- single
measurements, either
the only one--two known, the most reliable, or the brightest in a
series. Open symbols ---  averages for some period of time (Thum \et\ 1992;
1994a).
The observed values of flux for 2\al, 3\al, and 4\al\ were corrected
for extinction, in accordance with the Fig.~5 of Thompson \et\
(1977). Error bars reflecting internal convergency are not shown: they are
comparable to the symbol sizes for all observations. ``Error bars'' by H\al\
show the amplitude of variations observed by Greenstein (1973) during one
day (they were not intrinsic variations in \mwc-A --- see footnote
$^1\,$ in Brugel and Wallerstein, 1979).
Two solid lines give, at two extreme values of $N_e$, the flux {\it
ratios} predicted by the optically
thin LTE model (``Case B'', $T_e = 10,000\,$ K). The dashed line
takes into account free-free absorption at low frequencies. Vertical
position of all theoretical lines is arbitrary.
\ls
Fig.~2. Location, within the model disk, of the annuli of maximum
unsaturated amplification for $n\alpha\,$ lines. The values of $n\,$ are
indicated along the top ``border'' of the disk. The two ``borders''
give, in fact, only an idea of the disk's flare, showing the Log of
the scale hight (in $1.8\cdot 10^{10}\,$cm) as a function of radius
(in $1\cdot 10^{12}\,$cm). One quarter of the disk is only shown. The
width of every annulus (at half maximum of $k_{net}^{max}$) is
$\approx \pm 0.2\,$ in logarithm. Hence, in this logarithmic scale
all the annuli have approximately the same width, and it is seen that
overlapping of annuli should increase with increasing $n$.
\ls
Fig.~3. To the explanation of the closeness of excitation
temperatures (inversions) in \al-\ and \be-lines of close $n$.
\vfill\eject
VLADIMIR S. STRELNITSKI: Laboratory for Astrophysics, MRC 321,
National Air and Space Museum, Smithsonian Institution, Washington,
DC 20560.
{\it e-mail:} vladimir@wright.nasm.edu
\vskip6pt
HOWARD A. SMITH: Laboratory for Astrophysics, MRC 321,
National Air and Space Museum, Smithsonian Institution, Washington,
DC 20560.
{\it e-mail:} howard@wright.nasm.edu
\vskip6pt
VICTOR O. PONOMAREV: Astro-Space Center of P.N. Lebedev Physical
Institute, Leninsky prospect, 53, Moscow 117924. {\it e-mail:}
ponomarev@rasfian.serpukhov.su
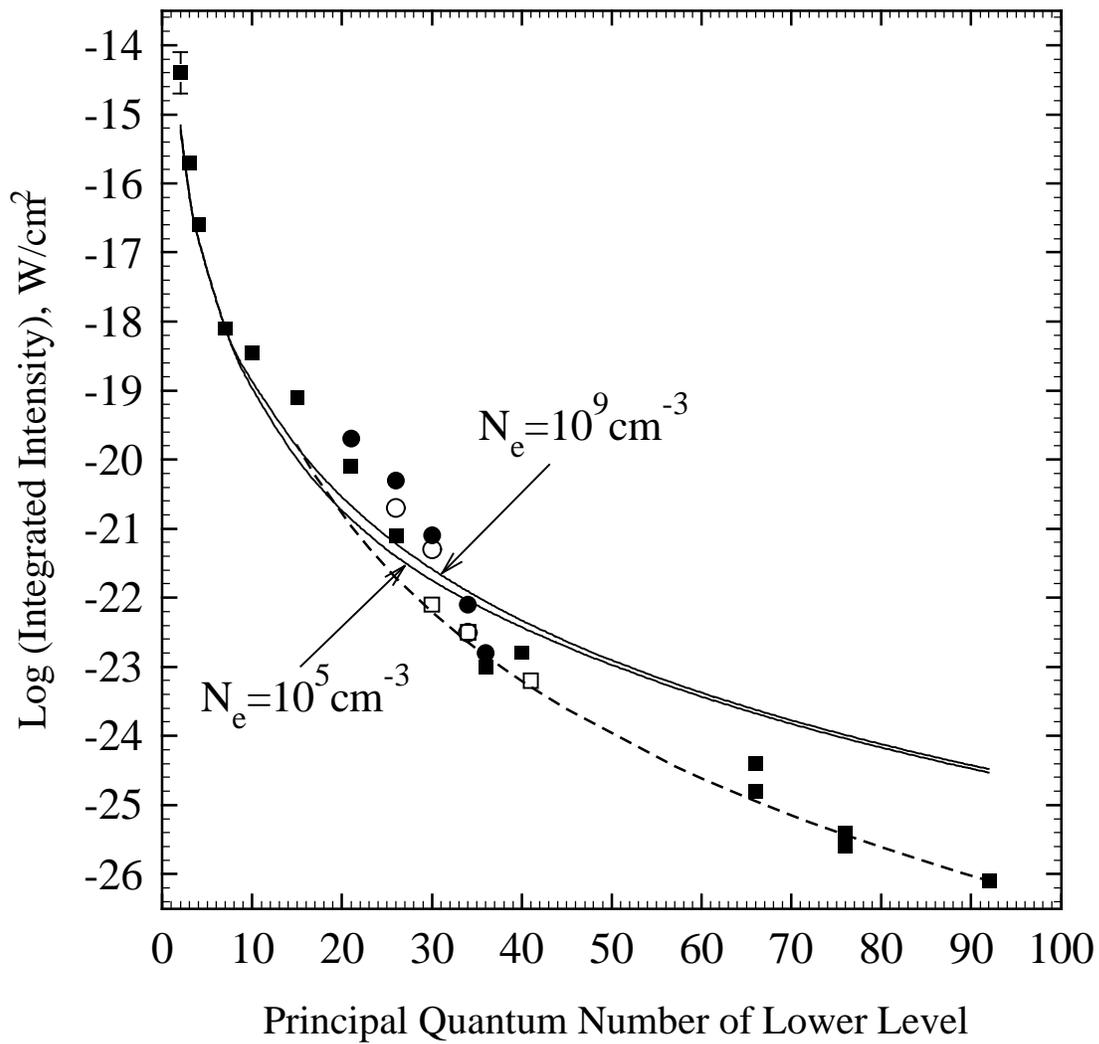
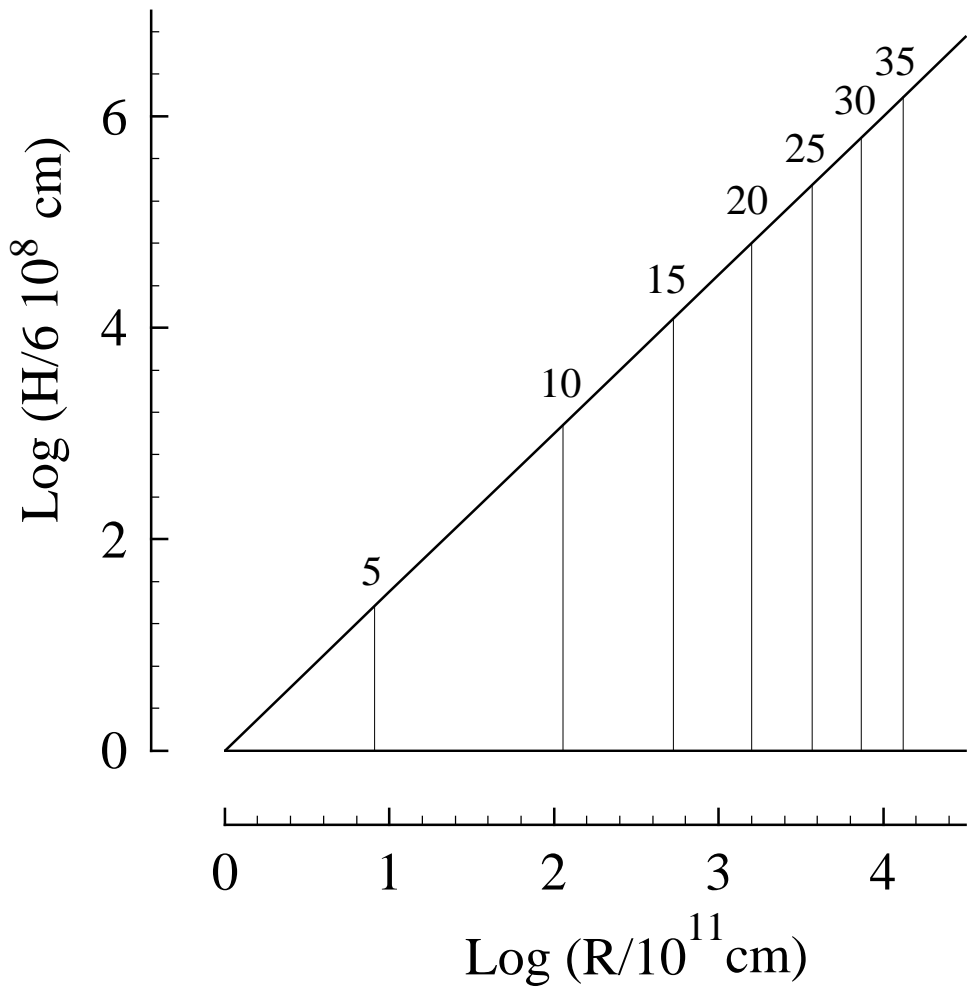
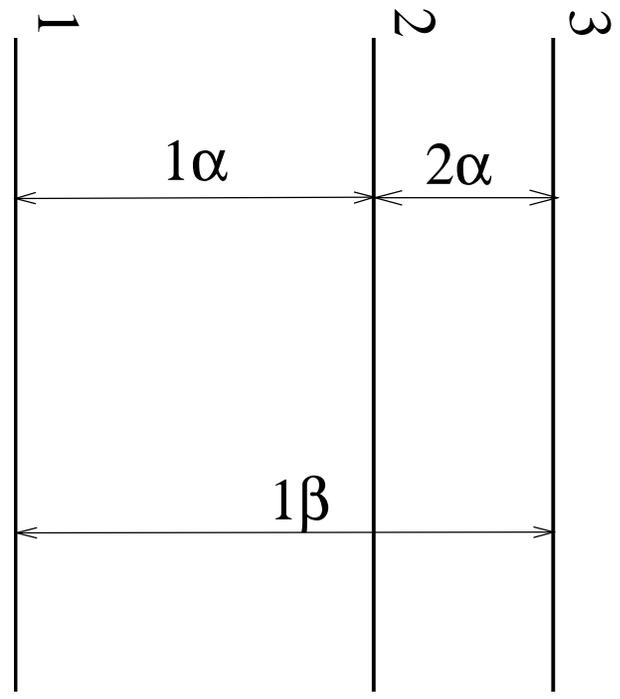
\end